\documentclass[secnumarabic,amssymb, nobibnotes, aps, prd, 11pt, abstract]{revtex4-2}
\usepackage{tocloft}
\usepackage{graphicx}
\usepackage{amsmath}
\usepackage{subcaption}
\usepackage{hyperref}
\hypersetup{colorlinks,linkcolor={red},citecolor={blue},urlcolor={red}}
\usepackage{setspace}
\usepackage{xcolor}
\pagecolor{white}

\usepackage{multirow}
\color{black}

\begin{document}

\singlespacing

	\title{Restricted Phase Space Thermodynamics of NED-AdS Black Holes }%

	\author{Mozib Bin Awal$^1$}
	
	\email{$mozibawal@gmail.com$}
	
	\author{Prabwal Phukon$^{1,2}$}
	\email{$prabwal@dibru.ac.in$}
	
	\affiliation{$^1$Department of Physics,Dibrugarh University, Dibrugarh,Assam,786004.\\$^2$Theoretical Physics Division, Centre for Atmospheric Studies, Dibrugarh University, Dibrugarh,Assam,786004.\\}

	\begin{abstract}
	
We study the Restricted Phase Space Thermodynamics (RPST) of magnetically charged Anti de Sitter (AdS) black holes sourced by nonlinear electrodynamics(NED). The first law and the corresponding Euler relation are examined using the scaling properties. While the mass is homogeneous in the first order, the intensive variables are observed to follow zeroth order homogeneity. We use numerical and graphical techniques to find the critical points of the various thermodynamic quantities. By utilizing the re-scaling properties of the equation of states, we study the thermodynamic processes using different pairs of variables. From our analysis, we infer that although the RPS thermodynamics of NED-AdS black hole resembles those of RN-AdS, Kerr-AdS, Kerr-Sen-Ads black holes in most of its aspects, hinting at a possible universality, there exists one particular $\mu-C$ process that differs in its behaviour from its counterparts in earlier reported works. 
\end{abstract}
	
	\maketitle
	
\section{Introduction}
Since the initial days of black hole thermodynamics\cite{1, area, Beken,Bard} AdS black holes have drawn considerable attention due to their interesting phase structures \cite{Kubiz,Hawking:1982dh,Cai}. Maladacena's proposal of AdS-CFT correspondence \cite{Maldacena} in 1997 has further fuelled this trend. In the last fifteen years or so, black hole thermodynamics has added a number of new formalisms into its domain. One such formalism that has gained prominence in recent years is the extended phase space thermodynamics \cite{Kastor, Dolan, Dolan2, Dolan3,Cai,Kubizna,Xu,Xu2,Zhang}. In this version of black hole thermodynamics, one defines two new conjugate thermodynamic variables, pressure and volume. The thermodynamic pressure, $P$ is defined in terms  negative cosmological constant $\Lambda$ as follows: $$P=-\Lambda/8\pi G$$

Application of this formalism to study AdS black holes have culminated in a number of novel phase behaviours in these black holes \cite{Kubiz,Xu,Xu2,Zhang,John,Xuahm}. A recent modified version of extended pahase space thermodynamics has been developed by Visser \cite{Visser} using the AdS/CFT correspondence.In this approach,the central charge and the chemical potential are included as new cinjugate pair $(\mu, C)$ in the first law. It must be noted that the inclusion of chemical potential and central charge as thermodynamically conjugate variables was explored earlier \cite{Kastor2014}. For example in \cite{Zhang2015,Karch,Maity,Wei}, such idea was used, but in these analysis the idea of the the central charge was vague \cite{Karch,Maity,Wei,Rafiee}. In Visser's formalism, he transformed the volume and pressure of the bulk AdS into that of the CFT side's with the help of the AdS/CFT correspondence, i.e. $\mathcal{V} \sim L^{d-2}$, with $\mathcal{V}$ denoting the CFT volume and $L$ is the usual AdS length. The corresponding pressure term, $\mathcal{P}$ can be determined using the Equation of State (EOS) of the CFT side, $E=(d-2)\mathcal{P}\mathcal{V}$, $d$ being the dimensions of the spacetime in the bulk. This is the main difference of Visser's formalism from the earlier ideas, he transformed $(P,V)$ variables in the bulk to $(\mathcal{P},\mathcal{V})$ in the boundary. With this, the first law can be written as, \begin{equation*}
dE=TdS-\mathcal{P}d\mathcal{V}+\tilde{\Phi}d\tilde{Q}+\Omega dJ+\mu dC
\end{equation*} This in turn is accompanied by an Euler-like relation\begin{equation}\label{eq1}
E=TS+\tilde{\Phi}\tilde{Q}+\Omega J+\mu C
\end{equation} where $\tilde{Q}$ and $\tilde{\Phi}$ are the rescaled electric charge and its conjugate potential. One of the issues of EPS formalism is that in this formalism we have to interpret the mass as enthalpy in contradiction to what the traditional black hole thermodynamics suggests- that the mass should be interpreted as internal energy. In Visser's formalism we get rid of this problem. His formalism allows us to again interpret the mass as internal energy. Visser's formalism is a complete understanding of the black hole thermodynamics in the bulk, as well as in the CFT boundary. In this framework one can describe the thermodynamics of a holographically dual conformal field theory in the boundary as well the thermodynamics of the bulk \cite{Rafiee,Cong}. Another problem with the EPS formalism is the absence of a homogeneity relation between the internal energy and intensive variables. Other than that, EPS formalism suffers from the issue of ``ensemble of theories''. Notably, this issue cannot be resolved in Visser's framework as well.
To avoid such issues, the authors in \cite{Zeyuan} proposed a new version of the Visser's proposed formalism. It has been called Restricted Phase Space (RPS) formalism. In \cite{Zeyuan} the authors studied the  RN-AdS black hole in the same formalism. This formalism omits the $(P,V)$ terms and hence is called Restricted Phase Space Thermodynamics (RPST). RPS formalism however allows us to vary $G$. Unlike varying $\Lambda$, varying $G$ does not change the geometry and so we do not have to face the ensemble of theories issues. As a continuation of \cite{Zeyuan}, the authors also studied the Kerr-AdS black hole system in \cite{Gao}. Following these, quite a few numbers of works have been done in the RPST formalism. Refs. \cite{Ali,Ladgham,Kong,Kong2,Du,Sadeghi,Sadeghi:2023dsg,Bai,Alip,Sadeg} are to cite a few.

In this paper we extend the investigation of thermodynamics within the RPS framework to NED-AdS black holes. We use numerical and graphical techniques to find the critical points of various thermodynamic quantities and analyse the thermodynamic processes using conjugate pairs of variables. 

The paper's organization is outlined as follows. In Section \ref{sec2}, we provide a brief overview of the NED-AdS black hole, derive several thermodynamic quantities, and demonstrate the validity of the first law and the Euler relation. In Section \ref{sec3}, we present an alternative expression for the black hole mass using intensive variables such as entropy $S$, CFT charge $C$, magnetic charge $Q$, and AdS length $l$. Section \ref{sec4} is dedicated to the examination of various types of thermodynamic processes. Lastly, Section \ref{sec5} offers a concise summary of our findings.

	 \section{NED-AdS black hole and the RPS formalism }\label{sec2}
Let us begin with the action of the NED-AdS black holes \cite{Kruglov:2022mde}, \begin{equation}\label{eq2}
I=\int d^4x \sqrt{-g} \left(\frac{R-2 \Lambda }{16 \pi  G}+\mathcal{F}( \mathcal{L})\right)
\end{equation} where $\Lambda=-3/l^2$ is the cosmological constant, $l$ is the AdS radius and $G$ is the Newton's constant. The lagrangian \cite{Bron} the NED-AdS black holes is written as \begin{equation}\label{eq3}
\mathcal{L}( \text{$\mathcal{F}$})=-\frac{\mathcal{F}}{4 \pi  \text{cosh}^2\left(a\sqrt[4]{2\mathcal{|F|}}\right)}
\end{equation} where $a$ is called the coupling constant and $\mathcal{F}=F^{\mu\nu}F_{\mu\nu}/4$ is the field invariant. We analyse the spacetime with spherical symmetry which has the line element squared, \begin{equation}\label{eq7}
ds^2=-f(r)dt^2+\frac{1}{f(r)}dr^2+r^2(d\theta^2+\sin^2\theta d\phi^2)
\end{equation}
We will exclusively examine magnetized black holes as being electrically charged black holes, for NED that has Maxwell weak-field limit, results in singularities \cite{Bron}. The metric function in equation (\ref{eq7}) can be expressed as, \begin{equation}\label{eq14}
f(r)=1-\frac{2MG}{r}+\frac{Q^2G}{br}\tanh\left(\frac{b}{r}\right)+\frac{r^2}{l^2}
\end{equation} Here, for simplicity, the parameter $b=a\sqrt{q}$ is used. The mass of the black hole can be found out in terms of horizon radius by putting $f(r_+)=0$. This gives, \begin{equation}\label{eq15}
M=\frac{Q^2 \tanh \left(\frac{b}{r_+}\right)}{2 b}+\frac{r_{+}^{3}}{2 G l^2}+\frac{r_+}{2 G}
\end{equation} Using the metric function, the Hawking temperature is calculated and it xomes out to be,  \begin{equation}\label{eq16}
T=\frac{l^2 r_{+}^{2}+3 r_{+}^{4}-G l^2 Q^2 \text{sech}^2\left(\frac{b}{r_+}\right)}{4 \pi  l^2 r_{+}^{3}}
\end{equation} Having the temperature and mass, the entropy of the black hole can be easily figured out which comes out to be a very simple expression,\begin{equation}\label{eq17}
S=\frac{\pi r_{+}^{2}}{G}
\end{equation}
Additionally, the magnetic potential $\Phi$ is given by, \begin{equation}
\Phi=\frac{q}{b}\tanh\left(\frac{b}{r_+}\right)
\end{equation}
Similar to the works in \cite{Zeyuan,Gao,Ali}, we introduce the additional pair of thermodynamic variables for the RPST formalism,\begin{equation}\label{eq18}
C=\frac{l^2}{G}, \quad \mu=\frac{M-TS-\hat{\Phi}\hat{Q}}{C}
\end{equation} In the above equation (\ref{eq18}), $\hat{Q}$ is termed as the rescaled electric charge and $\hat{\Phi}$ is it's corresponding rescaled electric potential. These two quantities are expressed as \begin{equation}\label{eq19}
\hat{Q}=\frac{Ql}{\sqrt{G}}, \quad \hat{\Phi}=\frac{\Phi\sqrt{G}}{l}
\end{equation} here, we wish to make a couple of comments about the chemical potential $\mu$. First, all the thermodynamic variables written above, always need to be positive, apart from the chemical potential $\mu$. Secondly, the chemical potential can be identified as an independent thermodynamic quantity via the of AdS/CFT dictionary \cite{Gib,Cham,Gibb}.

 As mentioned earlier, the RPST formalism is different from Visser's proposed formalism using AdS/CFT in that, in the former we keep the AdS length, $l$ fixed and so its variation vanishes. Taking this into account and utilising the thermodynamic quantities above, it is evident that the first law of thermodynamics is valid., \begin{equation}\label{eq20}
dM=TdS+\hat{\Phi}d\hat{Q}+\mu dC
\end{equation} which is clearly free from the $\mathcal{P}d\mathcal{V}$ term. Also, equation (\ref{eq18}) implies that the Euler-like relation is followed,\begin{equation}\label{eq21}
M=TS+\hat{\Phi}\hat{Q}+\mu C
\end{equation} Equation (\ref{eq20}) and (\ref{eq21}) serve as the cornerstone equations within the RPS formalism. Additionally, these equations may inherently be capable of determining the probable Hawking-Page-like phase transitions. As mentioned in \cite{Zeyuan,Gao,Ali}, we should keep in mind that even though the $(\mu, C)$ variable pair is taken from the dual CFT, $\mu$ can actually be perceived as the chemical potential and $C$ can be understood as the effective number $N_{\text{bulk}}$ of the microscopic degrees of freedom of black hole in the bulk. Alternatively, we may put forward new rules to the AdS/CFT dictionary, viz, $\mu_{\text{CFT}}=\mu_{\text{bulk}}$ and $C=N_{\text{bulk}}$. However, for ease of notation, we drop the subscripts and just keep the original $(\mu,C)$ instead of $(\mu_{\text{bulk}},N_{\text{bulk}})$, bearing in mind that we study the black hole's thermodynamics in the bulk and not that of the dual CFT. Another thing to draw our attention at is the fact that the $(\mu,C)$ variable pair in the bulk, when re-scaled as: $\mu\rightarrow\lambda^{-1}\mu$ and $C\rightarrow\lambda C$, $\lambda$ being an arbitrary non-zero constant, keeps the first law and the Euler-like relation stays intact.

\section{Equation of States and Homogeneity}\label{sec3}
In this section we write the expressions of the thermodynamic quantities such as mass, temperature etc. in terms of the extensive variables $S, \hat{Q}$ and $C$. For that, we first express $G$ and $r_+$ in terms of $C$, $S$. This can be very easily done using equations (\ref{eq17}) and (\ref{eq18}). So, we have, \begin{equation}\label{eq22}
G=\frac{l^2}{C}, \quad r_+=\left(\frac{SG}{\pi}\right)^{1/2}
\end{equation} Using these two equations and the rescaled electric charge and potential in equation (\ref{eq19}), we write the mass $M$ in equation (\ref{eq15}) as,\begin{equation}\label{eq23}
M(S,\hat{Q},C)=\frac{1}{2} \left[\frac{\hat{Q}^2 \tanh \left(\frac{\sqrt{\pi } b}{\sqrt{\frac{l^2 S}{C}}}\right)}{b C}+\frac{(\pi  C+S) \sqrt{\frac{l^2 S}{C}}}{\pi ^{3/2} l^2}\right]
\end{equation} With the expression of mass in hand, and with the aid of the first law (\ref{eq20}), we can straight away write the Equation of States (EOS) as,\begin{equation}\label{eq24}
T=\left(\frac{\partial M}{\partial S}\right)_{\hat{Q},C}=\frac{l^2 \left[S (\pi  C+3 S)-\pi ^2 \hat{Q}^2 \text{sech}^2\left(\frac{\sqrt{\pi } b}{\sqrt{\frac{l^2 S}{C}}}\right)\right]}{4 \pi ^{3/2} C^2 \left(\frac{l^2 S}{C}\right)^{3/2}}
\end{equation}
\begin{equation}\label{eq25}
\hat{\Phi}=\left(\frac{\partial M}{\partial \hat{Q}}\right)_{S,C}=\frac{1}{bC}\left[\hat{Q} \tanh \left(\frac{\sqrt{\pi } b}{\sqrt{\frac{l^2 S}{C}}}\right)\right]
\end{equation}
\begin{equation}\label{eq26}
\mu =\left(\frac{\partial M}{\partial C}\right)_{S,\hat{Q}}=\frac{\hat{Q}^2 \left[\pi ^2 b \text{sech}^2\left(\frac{\sqrt{\pi } b}{\sqrt{\frac{l^2 S}{C}}}\right)-\frac{2 \pi ^{3/2} l \sqrt{S} \tanh \left(\frac{\sqrt{\pi } b \sqrt{C}}{l \sqrt{S}}\right)}{\sqrt{C}}\right]+b S (\pi  C-S)}{4 \pi ^{3/2} b C^2 \sqrt{\frac{l^2 S}{C}}}
\end{equation}
Therefore, we have expressed the intensive variables as functions of the extensive variables $S$, $\hat{Q}$ and $C$. We can observe that, with the rescaling $S\rightarrow \lambda S$, $\hat{Q}\rightarrow \lambda \hat{Q}$ and $C\rightarrow \lambda C$  the mass $M$ in equation (\ref{eq23}) rescales as $M\rightarrow \lambda M$. However, $T$,$\hat{\Phi}$ and $\mu$ in equation (\ref{eq24}), (\ref{eq25}) and (\ref{eq26}) does not get rescaled the same way proving the homogeneity of $M$ in the first order and the homogeneity of $T$,$\hat{\Phi}$ and $\mu$ in $S, \hat{Q}, C$ in zeroth order. At this point, we may remind ourselves that mathematically, the functions which are homogeneous in the zeroth order are by definition intensive. Considering all these arguments and the first law (\ref{eq20}) and it's corresponding Euler-like relation (\ref{eq21}), we can write the Gibbs-Duhem relation as, \begin{equation*}
d\mu=-\mathcal{S}dT-\hat{\mathcal{Q}}d\hat{\Phi}
\end{equation*}with $\mathcal{S}=S/C$ and $\hat{\mathcal{Q}}=Q/C$ are homogeneous functions of $S, \hat{Q}$ and $C$ of zeroth order. Any standard thermodynamic system has to obey the first law of black hole mechanics, the Euler-like relations and the Gibbs-Duhem relation. Moreover, the homogeneity properties of the intensive variables including the internal energy must also be obeyed. RPST formalism satisfies all these relations and homogeneity properties which were absent in the previous formalisms of black hole thermodynamics.

It is evident that equations (\ref{eq24}) to (\ref{eq26}) give three algebraic relations for six parameters in total. As a result, a black hole macrostate can be described by three of the six thermodynamic parameters $(T, S)$, $(\hat{\Phi}, \hat{Q})$ and $(\mu, C)$. So, variation of any three variables can characterise a macroscopic process and to study the thermodynamic behaviour we have to vary atleast three variables. One last comment before going into the analysis of the thermodynamic processes in the next section - we notice that $\hat{Q}$ always appears as $\hat{Q}^2$ in equations (\ref{eq23}), (\ref{eq24}) and (\ref{eq26}). However from equation (\ref{eq25}), it can be inferred that $\hat{\Phi}$ and $\hat{Q}$ must always have identical signs. So, we may conclude that considering only the cases with $\hat{Q}\geq 0$ will be sufficient.

\section{Thermodynamic Processes}\label{sec4}
The goal of this section is to analyse the various thermodynamic processes of the NED-AdS black hole in RPS formalism. In the previous section we discussed from the equation of states that the intensive variables are functions of three extensive variables. So, it will be a very complicated task to analyse all the allowed thermodynamic processes. Therefore we will only focus on some of the special kinds of processes by keeping two of independent variables constant. We will consider simple, conventional processes having only one pair of canonically conjugate variables where one is intensive and the other is extensive. However, in the following sections, we will see that even the simple processes can lead to complicated results and one must resort to alternative methods such as numerical or graphical techniques for the solution of the equations. The Reissner-Nordstr\"{o}m-AdS black hole is analysed in Ref \cite{Zeyuan}, and in this case one gets the exact analysis. It is required to take approximations in the slow rotating limits to get the analytic results for Kerr-AdS black holes. For Kerr-Sen-Ads black holes \cite{Ali}, the authors used numerical techniques for the analysis. In our case also we adopt some graphical and numerical techniques to study the thermodynamic processes.

\subsection{Critical Points and $T-S$ processes}The study of the $T-S$ processes is essential in order to decode the information about first order phase transitions. These transitions occur below the critical points and at the critical point, they become second order. This is true for both the EPS and RPS fromalism. Obviously, the $T-S$ curves has to be analysed at fixed $\hat{Q}$ (or $J$ in case of Kerr or Kerr-Sen-AdS black holes).

Now, we aim to find the critical points of the $T-S$ curve at constant $\hat{Q}$. In order to do that we have to solve for an inflection point in the $T-S$ curve. This can be done by solving the two equations given below \begin{equation}\label{eq27}
\left(\frac{\partial T}{\partial S}\right)_{\hat{Q},C}=0, \quad \left(\frac{\partial^2 T}{\partial S^2}\right)_{\hat{Q},C}=0
\end{equation} To solve these, we employ the equation of state (\ref{eq24}). This set of equation becomes extremely complicated and is not possible to find an analytic solution. Therefore, we use graphical and numerical technique. In doing so, we will get the approximate critical points.

We know, by dimensional analysis, that we must have, \begin{equation}\label{eq28}
S_c=p_1C, \quad \hat{Q}_c=p_2C
\end{equation} where $p_1$ and $p_2$ are some coefficients. We need to keep in mind such scalings while employing numerical or graphical techniques. After solving we get the critical values as \begin{equation*}
S_c\approx 0.4947C, \quad \hat{Q}_c\approx 0.1754C
\end{equation*}Accordingly, the critical temperature and the Helmholtz free energy, $F=M-TS$ comes out to be \begin{equation*}
T_c\approx 0.2584l^{-l}, \quad F_c\approx 0.1398l^{-l}C
\end{equation*} 

\begin{figure}[h]
\centering
\begin{subfigure}{.5\textwidth}
  \centering
  \includegraphics[height=6.3cm,width=.95\linewidth]{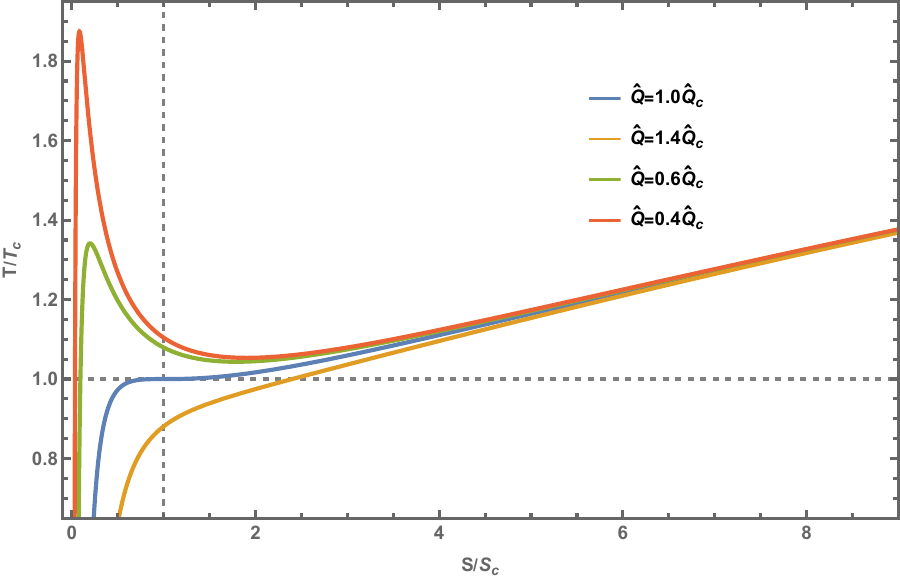}
  \caption{}
\end{subfigure}%
\begin{subfigure}{.5\textwidth}
  \centering
  \includegraphics[height=6.5cm,width=.95\linewidth]{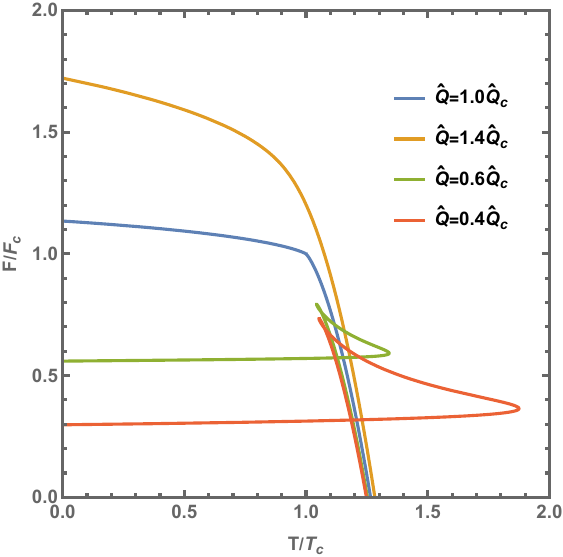}
  \caption{}
\end{subfigure}
\caption{(a) $T-S$ and (b) $F-T$ curves at fixed $\hat{Q}>0$ (iso-\textit{e}-charge)}
\label{ts ft}
\end{figure}

At these critical points, the black hole undergoes a phase transition. For instance, in Figure \ref{ts ft} we observe that above the critical point $T_c$ there are three black hole phases and below the critical point only one phase exists. This is analogous to the liquid gas phase transition in classical thermodynamics. The specific heat at constant value of charge is given by \begin{equation*}
\mathbf{C}=T\frac{dS}{dT}
\end{equation*}Which comes out to be \begin{equation}\label{sp}
\mathbf{C}=\frac{2 S \sqrt{\frac{l^2 S}{C}} \left[S (\pi  C+3 S)-\pi ^2 \hat{Q}^2 \text{sech}^2\left(\frac{\sqrt{\pi } b}{\sqrt{\frac{l^2 S}{C}}}\right)\right]}{S (3 S-\pi  C) \sqrt{\frac{l^2 S}{C}}-\pi ^2 \hat{Q}^2 \left(2 \sqrt{\pi } b \tanh \left(\frac{\sqrt{\pi } b}{\sqrt{\frac{l^2 S}{C}}}\right)-3 \sqrt{\frac{l^2 S}{C}}\right) \text{sech}^2\left(\frac{\sqrt{\pi } b}{\sqrt{\frac{l^2 S}{C}}}\right)}
\end{equation} The specific heat should diverge at the critical points. Putting the calculated values of the critical points in the denominator of equation \ref{sp} we find it becomes approximately zero (approximately because the critical values are not exact but numerically calculated. It comes out of the order of $10^{-6}$).

Some approaches have been suggested to probe black hole phase trnasitions by examining Quasinormal Modes (QMNs) \cite{qnmpt1,qnmpt2,qnmpt3}. QNMs characterise the perturbations in the surrounding geometry of a black hole. A QMN frequency has both real and imaginary part describing oscillation and damping time respectively, hence they are referred to as the characteristic sounds of the black hole. These are believed to be potentially detected in the gravitational wave detector \cite{qnmgw1,qnmgw2}. QNMs and thermodynamic phase transition has been studied for a number of black holes and it has been found that the QNMs show contrasting behaviour in different black hole phases. Specifically, the modes completely change when we go from a small black hole phase to a large black hole phase. Hence it is conclusive that the information of phase transition could be extracted from the gravitational waves.

With the critical points, we can now re-write the equation of state (\ref{eq24}) and free energy in terms of the what has been called in literature as the `` reduced parameters ",\begin{equation*}
s=\frac{S}{S_c},\quad \tau=\frac{T}{T_c}, \quad q=\frac{\hat{Q}}{\hat{Q}_c}
\end{equation*} We can now write the expressions of temperature and free energy as in terms of the reduced parameters as, \begin{equation}\label{eq29}
\tau(s,q)=\frac{(0.376107 s+0.747786) s-0.123947 q^2 \text{sech}^2\left(\frac{0.00244227}{\sqrt{s}}\right)}{s^{3/2}}
\end{equation}
\begin{multline}\label{eq30}
f(s,q)= \frac{7.15308 s (0.0992056\, -0.0156217 s)+0.0193822 q^2 \text{sech}^2\left(\frac{0.252002}{\sqrt{s}}\right)}{\sqrt{s}}\\ +\frac{0.153826 q^2 \sqrt{s} \tanh \left(\frac{0.252002}{\sqrt{s}}\right)}{\sqrt{s}}
\end{multline}
We need to mention here that we have used $b/l=0.1$ while writing these equations. From the above equations, it is evident that when written in terms of the reduced parameters, in the equation of states and the free energy, there are no dependence of central charge $C$. So, black holes, even with different central charges will show the same thermodynamic behaviour. This can be seen in ordinary thermodynamics as well where instead of central, the number of particles has to be considered. This is the so-called \textit{the law of corresponding states}.

The plots of $T/T_c-S/S_c$ and $F/F_c-T/T_c$ for iso-\textit{e}-charge processes are given in Figure \ref{ts ft}. We can observe from the structure of the $T/T_c-S/S_c$ curves that below the critical point where $\hat{Q}<\hat{Q}_c$, there is an oscillatory phase. At $T>T_c$, we can observe that a black hole has three states with the same charge $\hat{Q}$ and central charge $C$. However their entropies are different at some temperature $T=T_0$. This is the characteristic of a phase transition of first-order. Correspondingly, the $F/F_c-T/T_c$ curves show a swallowtail behaviour indicating a Van der Waals like liquid-gas phase transition provided $0<\hat{Q}<\hat{Q}_c$. However when $\hat{Q}>\hat{Q_c}$, the $F/F_c-T/T_c$ curves show continuous behaviour. From Figure \ref{ts ft}(a) we can conclude that at the Hawking temperatures above the critical value $T_c$, there are three black hole states with same temperature, central charge and Q among which the small and large black hole states are stable. These two states coexist along with the metastable state when the first order phase transition occurs. However, when the Hawking temperature equals the critical value of the temperature, i.e. $T=T_c$, we observe that the three black hole phases merge into a single one. Similar phenomenon has been seen in RN-AdS, Kerr-AdS and Kerr-Sen-Ads black holes.

We now consider the processes where we keep $\hat{\Phi}$, $C$ fixed and plot the $T-S$ curves. They may be called isovoltage processes. We start by replacing $\hat{Q}$ in equation (\ref{eq24}) using equation (\ref{eq25}). We obtain, \begin{equation*}
T_{\hat{\Phi}}=\frac{l^2 \left[S (\pi  C+3 S)-\pi ^2 b^2 C^2 \hat{\Phi}^2 \text{csch}^2\left(\frac{\sqrt{\pi } b}{\sqrt{\frac{l^2 S}{C}}}\right)\right]}{4 \pi ^{3/2} C^2 \left(\frac{l^2 S}{C}\right)^{3/2}}
\end{equation*} We have a bound for $\hat{\Phi}$ here in order for $T$ to be positive, which is, \begin{equation*}
\hat{\Phi} \leq \frac{\sqrt{S} \sqrt{\pi  C+3 S} \sinh \left(\frac{\sqrt{\pi } b}{\sqrt{\frac{l^2 S}{C}}}\right)}{\pi  b C}
\end{equation*} We observe that there are no inflection points on the $T-S$ curves but they all have an extremum point. This point is the minimum value of $T$ located at the corresponding value of entropy $S=S_{min}$. For NED-AdS black hole, we cannot find the minimum temperature analytically. However, by plotting $T(S,\hat{\Phi},C)$ against $S$, we can find $T_{min}$ located at $S=S_{min}$. We can rexpress the temperature and free energy as functions of reduced parameters \begin{equation*}
\hat{s}=\frac{S}{S_{min}}, \quad \hat{\tau}=\frac{T}{T_{min}}, \quad \hat{f}=\frac{F}{F_{min}}
\end{equation*} as $\hat{\tau}=\tau(\hat{s},\hat{Q},C)$ and $\hat{f}=f(\hat{s},\hat{Q},C)$

\begin{figure}[h]
\centering
\begin{subfigure}{.5\textwidth}
  \centering
  \includegraphics[height=6.3cm,width=.95\linewidth]{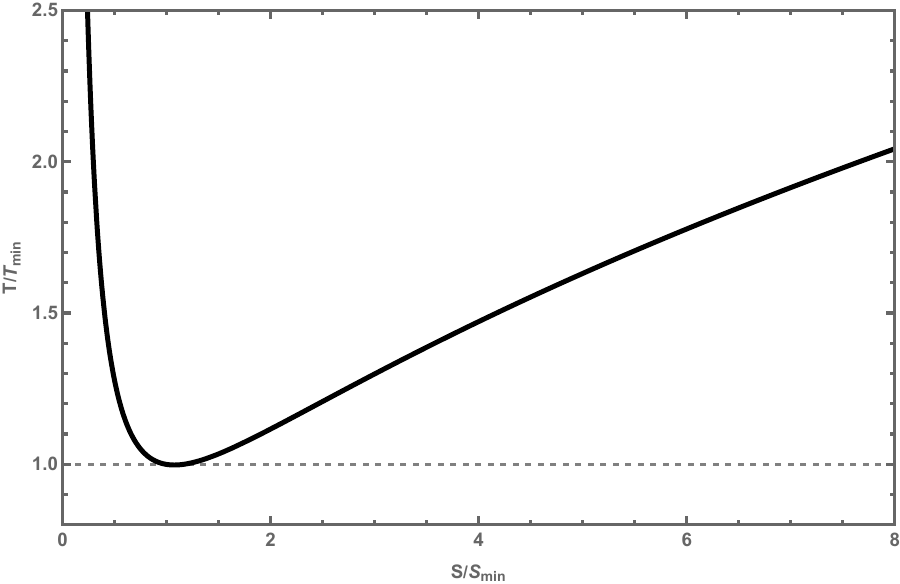}
  \caption{}
\end{subfigure}%
\begin{subfigure}{.5\textwidth}
  \centering
  \includegraphics[height=6.3cm,width=.95\linewidth]{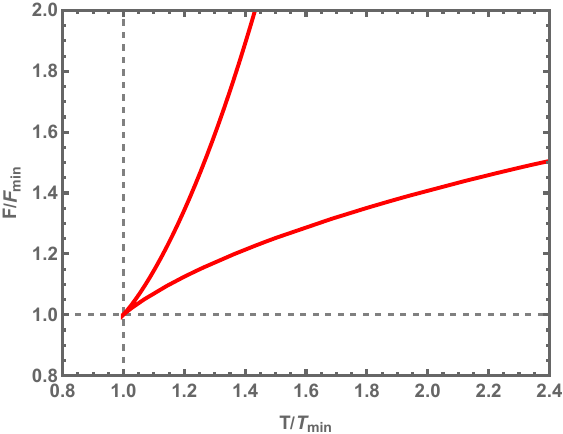}
  \caption{}
\end{subfigure}
\caption{(a) $T-S$ and (b) $F-T$ curves at fixed $\hat{\Phi}$ (isovoltage)}
\label{ts ft min}
\end{figure} We provide the plots of $T/T_{min}-S/S_{min}$ and $F/F_{min}-T/T_{min}$ in Figure \ref{ts ft min}. We have to mention here that although it is not possible to find the minimum temperature and entropy analytically, we observe that for all the different values of $\hat{\Phi}$, the respective curves scale down to a single curve after using the reduced parameters. In this case also, we have used $b/l=0.1$. It is evident from the Figure \ref{ts ft min} that for temperatures $T>T_{min}$, two black hole states will exist having same $T$, $C$ and $\hat{\Phi}$ but with different entropies where, the black hole state with the smaller entropy will correspond to the unstable state and the one with the larger entropy corresponds to the stable state. We also observe that the plot of free energy with temperature on the right panel does not go to zero anywhere implying there are no Hawking-Page transition  occuring here. We will discuss more on the Hawking-Page transitions later.

We may also want to see the $\hat{\Phi}-\hat{Q}$ processes. However, from equation (\ref{eq25}) we can observe that $\hat{\Phi}$ is just proportional to $\hat{Q}$, so, no inflection or extremum point will exist on the $\hat{\Phi}-\hat{Q}$ curves.

\subsection{$\mu-C$ processes}
In this section we aim to explore the $\mu-C$ curves i.e. here we allow the central charge $C$ to vary. Here $\mu$ has to be understood as the Gibbs free energy per unit central charge. We start with the expression for $\mu$ in equation (\ref{eq26}). From this equation we find that there lies only one extremum on each of the $\mu-C$ curves at fixed $(S, \hat{Q})$ corresponding to a maximum of $\mu$ at $C=C_{max}$. It is not possible to solve for the maximum values of $\mu$ and $C$ for the NED-AdS black hole system. However, as employed in the previous case, we shall use the reduced parameters,\begin{equation*}
c=\frac{C}{C_{max}}, \quad m=\frac{\mu}{\mu_{max}}
\end{equation*} In doing so, we observe that all the $\mu-C$ curves scale down to a single one. Note that, for any given values of $S$ and $\hat{Q}$, the value of $\mu_{max}$ always comes out as positive. 
 \begin{figure}[h]
    \centering
    \includegraphics[height=6.4cm,width=.54\linewidth]{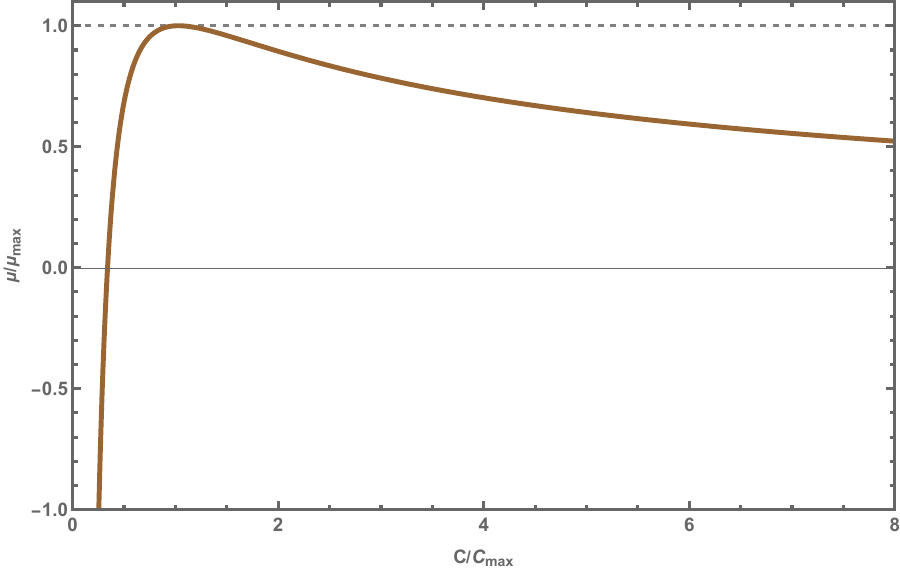}
    \caption{$\mu-C$ curve in the adiabatic iso-\textit{e}-charge process}
    \label{mu c}
\end{figure} Figure \ref{mu c} shows the $\mu/\mu_{max}-C/C_{max}$ curves for NED-AdS black hole. In this Figure, we can see that the chemical potential goes to zero for a certain central charge. Such zeroes correspond to the Hawking-Page phase transition. Besides that, it is also important to mention that similar behaviour has been observed in RN-AdS, Kerr-AdS and Kerr-Sen-AdS black hole. This may not be just a coincidence but may indicate that there lies some universality behind the $\mu-C$ processes at fixed $J$ (in case of Kerr-AdS), $\hat{Q}$ (in case of RN-AdS and NED-AdS) and $J, \hat{Q}$ (in case of Kerr-Sen-AdS).

We can also study the $\mu-C$ processes at constant $S, \hat{\Phi}$. In order to do that we replace $\hat{Q}$ in equation (\ref{eq26}) using equation (\ref{eq25}). We refrain ourselves from writing the cumbersome analytical expression. Howvever, using similar techniques used in the previous process, we plot the $\mu-C$ curve. Throughout this analysis we have to be careful while fixing $\hat{\Phi}$ because of it's bound. 
\begin{figure}[h]
\centering
\begin{subfigure}{.5\textwidth}
  \centering
  \includegraphics[height=6.3cm,width=.95\linewidth]{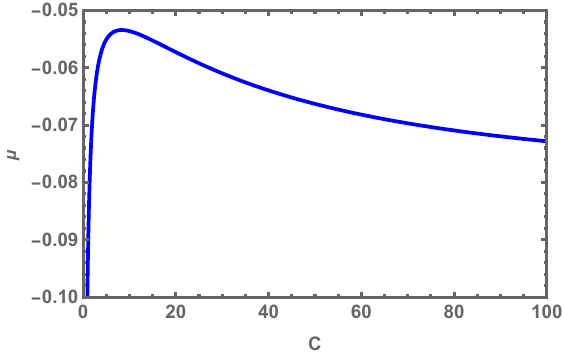}
  \caption{}
\end{subfigure}%
\begin{subfigure}{.5\textwidth}
  \centering
  \includegraphics[height=6.3cm,width=.95\linewidth]{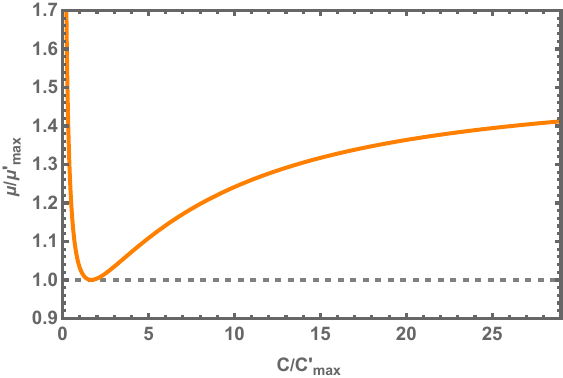}
  \caption{}
\end{subfigure}
\caption{$\mu-C$ process at fixed $\Phi$.}
\label{mu c fix phi}
\end{figure}
In Figure \ref{mu c fix phi}, we provide the plot of $\mu$ with $C$ at fixed $S, \hat{\Phi}$. It is found that just like the previous case, $\mu-C$ curve at constant $S, \hat{\Phi}$ also has a maximum but this time the maximum is not always positive, in fact, the maximum is always negative when $\hat{\Phi}$ is chosen according to it's bound. In the right panel we plot $\mu/\mu'_{max}-C/C'_{max}$ at fixed $\hat{\Phi}$. Since $\mu'_{max}$ is negative, we see the plot goes upside down. Clearly, nowhere does the chemical potential $\mu$ go to zero, implying there are no Hawking-Page transition. This is consistent with the result in the preceding section where we saw that the free energy in Figure \ref{ts ft min} never went to zero. This result is in contrast with the RN-AdS case, where two $\mu-C$ process were analysed at fixed $\hat{Q}$ and fixed $\hat{\Phi}$, and both of them were exactly the same.

\subsection{$\mu-T$ process}
\begin{figure}[h]
    \centering
    \includegraphics[height=6.2cm,width=.52\linewidth]{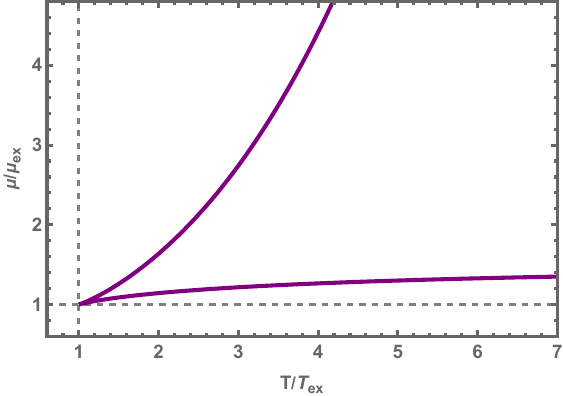}
    \caption{$\mu-T$ at fixed $\hat{\Phi}$}
    \label{mu T}
\end{figure}

For a better understanding of the Hawking-Page transition, we may also check the $\mu-T$ process at fixed $S, \hat{\Phi}$. $T$, as we have discussed earlier, has an extremum which corresponds to a minimum. The extremum of $\mu$ on the other hand corresponds to a maximum. $\mu$ is parametrically plotted against $T$. Figure \ref{mu T} shows the plot of $\mu/\mu_{ex}-T/T_{ex}$. It is again observed that the chemical potential nowhere becomes zero indicating the absence of any Hawking-Page transition. This confirms our analysis of the $\mu-C$ process at fixed $S, \hat{\Phi}$ in the previous section. Hence we may conclude that the universality of the $\mu-C$ processes does not apply to isovoltage processes. The $\mu-C$ processes, as it seems are not understood very well and needs further studies.

\section{Conclusions}\label{sec5}
We investigated the thermodynamics of NED-AdS black holes within the RPS framework. We studied the thermodynamics with the introduction of a new pair of thermodynamics variables, $(\mu, C)$ and keeping the AdS radius fixed. In this context, the Newton's constant is treated as a variable. In the RPS formalism, we observed that the first law aligns with the conventional description of ordinary thermodynamics, where the mass can be thought of as the internal energy. The Euler relation and the Gibbs-Duhem relation also hold perfectly. We then calculated the approximate critical values of temperature, free energy, entropy and the rescaled electric charge. Critical points are also found to occur with the usual Maxwell theory. The analysis is done using the RN-AdS black hole system \cite{Zeyuan}. However, we have observed that the numerical values of the critical points are totally different in the case when a black hole is sourced by non linear electrodynamics. Hence, we may conclude that the introduction of non linear electrodynamics alters the critical points. These critical points are associated with changes in the behaviour of the black hole, such as transitions between different thermodynamic phases (e.g., small and large black hole phases). So, addition of non linear electrodynamics should affect those behaviour. In this work we have demonstrated how some of the thermodynamic processes show different behaviour from that of the usual maxwell theory, while some processes remain the same.  Using the critical points, we studied the iso-\textit{e}-charge processes from the plots of temperature with entropy. It shows a first order phase transition reminiscent of Van der Waals behaviour for some non-vanishing subcritical value, $0<\hat{Q}<\hat{Q_c}$. The isovoltage process on the $T-S$ plane contains a single minimum and the free energy does not go to zero anywhere suggesting that a Hawking-Page transition is not present in this case. The constant $(S, \hat{Q})$ process have a single maximum on the $\mu-C$ plane and there exists one Hawking-Page transition point. While, no Hawking-Page points are observed in the fixed $(S, \hat{\Phi})$ process, which is confirmed by the $\mu-T$ process. This is in contrast with the universality proposed for the $\mu-C$ process in the existing literature \cite{Zeyuan,Gao,Ali}. The $\mu-C$ processes need further exploration for their full understanding. We plan to do so in our future works.

\end{document}